# Learning with Learned Loss Function: Speech Enhancement with Quality-Net to Improve Perceptual Evaluation of Speech Quality

*Szu-Wei Fu, Chien-Feng Liao, Yu Tsao*

*Abstract*—Utilizing a human-perception-related objective function to train a speech enhancement model has become a popular topic recently. The main reason is that the conventional mean squared error (MSE) loss cannot represent auditory perception well. One of the typical human-perception-related metrics, which is the perceptual evaluation of speech quality (PESQ), has been proven to provide a high correlation to the quality scores rated by humans. Owing to its complex and non-differentiable properties, however, the PESQ function may not be used to optimize speech enhancement models directly. In this study, we propose optimizing the enhancement model with an approximated PESQ function, which is differentiable and learned from the training data. The experimental results show that the learned surrogate function can guide the enhancement model to further boost the PESQ score (increase of 0.18 points compared to the results trained with MSE loss) and maintain the speech intelligibility.

## I. INTRODUCTION

IN recent years, various deep-learning-based models have been adopted for speech enhancement [1-14]. As compared to traditional methods, deep models have demonstrated notable improvements, especially under challenging test conditions (non-stationary noise and low signal-to-noise ratio). Despite the current success demonstrated by the deep-learning-based models, there are potential directions for further improvements.

This paragraph of the first footnote will contain the date on which you submitted your paper for review.

Szu-Wei Fu is with Department of Computer Science and Information Engineering, National Taiwan University, Taipei, Taiwan and Research Center for Information Technology Innovation (CITI) at Academia Sinica, Taipei, Taiwan. (e-mails: d04922007@ntu.edu.tw).

Chien-Feng Liao is with College of Electrical Engineering and Computer Science, National Taiwan University, Taipei, Taiwan and Research Center for Information Technology Innovation (CITI) at Academia Sinica, Taipei, Taiwan. (e-mails: r06946002@ntu.edu.tw).

Yu Tsao is with the Research Center for Information Technology Innovation (CITI) at Academia Sinica, Taipei, Taiwan (e-mails: yu.tsao @citi.sinica.edu.tw ).

One of the directions is to adopt a better objective function to train the models. Traditionally, the mean squared error (MSE) criterion is used as the objective function for optimizing the model parameters. However, the MSE scores may not indicate human auditory perception well. In fact, several researches pointed out that a processed speech with a small MSE score (compared to its clean counterpart), does not guarantee that it has high quality and intelligibility scores [15, 16]. Among the human-perception-related objective metrics, the perceptual evaluation of speech quality (PESQ) [17] and short-time objective intelligibility (STOI) [18] are two popular functions for evaluating speech quality and intelligibility, respectively. Therefore, optimizing the enhancement models directly using these two functions is a reasonable direction.

Several studies [15, 16, 19-24] have focused on STOI score optimization to improve speech intelligibility. Our previous study [15], for the first time, proposed optimizing the STOI score directly without any approximation in an utterance-based enhancement manner. The experimental results show that by combining STOI with MSE as an objective function, the speech intelligibility can be increased, which has been verified by a listening test. In addition, the recognition accuracy of enhanced speech tested on automatic speech recognition (ASR) can also be improved. The PESQ scores, however, cannot be increased by maximizing the STOI score as reported in [15].

Few [16, 19, 25, 26] have considered PESQ function as an objective function, because it is non-fully differentiable and significantly more complex compared to STOI. Reinforcement learning (RL) techniques such as deep Q-network and policy gradient were employed to solve the non-differentiable problem, as [25] and [16], respectively. Zhang *et al.* [19] applied direction sampling to implement approximate gradient descent. For the three works above, the original PESQ function was used while a different learning process was performed to optimize the model parameters. Meanwhile, a new PESQ-inspired objective function that considered symmetrical and asymmetrical disturbances of speech signals was derived in [26]. The experimental results confirmed that based on the PESQ-inspired objective function, the enhanced speech achieved higher PESQ scores as compared with the MSE-based one.

In this study, we propose to maximize the PESQ score of the enhanced speech without knowing any computation details of the function. Our basic idea is simple: As a deep learning model is a powerful mapping function, an approximated PESQ func-tion can be learned as an end-to-end model. Our previous paper [27] indicated that the model, termed Quality-Net, did not require clean references when computing scores (thus regarded as a non-intrusive quality estimation model) and could yield a high correlation to the PESQ function. In this paper, Quality-Net is concatenated after an enhancement model and served as an objective function. To maximize the PESQ score, we simply fixed the weights in Quality-Net and updated the weights in the enhancement model, so that the estimated quality score can be increased. Unlike the previous frame-based methods [16, 25, 26], our method is utterance-based, similar to the calculation of the PESQ. Our experimental results indicate that the gradients provided by Quality-Net can increase the PESQ scores of enhanced speech rapidly. In addition, a significantly higher score can be obtained as compared to the one given by the MSE-based loss function.

## II. PESQ Score Maximization

Because the PESQ function is highly complex and non-fully differentiable (the gradient cannot be back-propagated), it is difficult to directly optimize it as a training objective function of deep-learning-based speech enhancement models. Therefore, we attempt to maximize the PESQ score of enhanced speech by applying a PESQ-approximated function as the loss function. This surrogate is also a deep model learned from training data pairs of ([degraded speech, clean speech], PESQ score), where the bracketed terms represent the concatenation. We herein denoted this surrogate Quality-Net, the same as in our previous study [27]. The magnitude spectrogram is adopted as the input features. Therefore, after reading the whole spectrogram, Quality-Net can predict a score for speech quality. Notably, the Quality-Net used in this study differs from the previous one [27] in the following aspects: First, because of the different goals in these two studies, the Quality-Net used in this study is an intrusive estimation model (implying that a clean reference is required). Next, we replace the bidirectional long short-term memory (BLSTM) structure by a convolutional neural network (CNN) for optimization reasons (empirically, we found that by using CNN, gradients can more easily back-propagate to the previous enhancement model). Because the input of Quality-Net is magnitude spectrogram, it can be combined easily with a speech enhancement model whose outputs are also magnitude spectrogram. In the following, we introduce the two steps of the proposed framework.

### A. Training of Quality-Net

To approximate the PESQ function by Quality-Net, the output scores of these two functions should be as close as possible when they have the same inputs. Therefore, we first calculated the PESQ scores of the training data; subsequently, Quality-Net is trained with the MSE loss to minimize the difference between the estimated and true scores. As our framework performs on the utterance level (variable size), we apply the global average operation in Quality-Net to handle the limitations that conventional CNNs (ending with fully connected layers) can only predict the scores with fixed-size inputs.

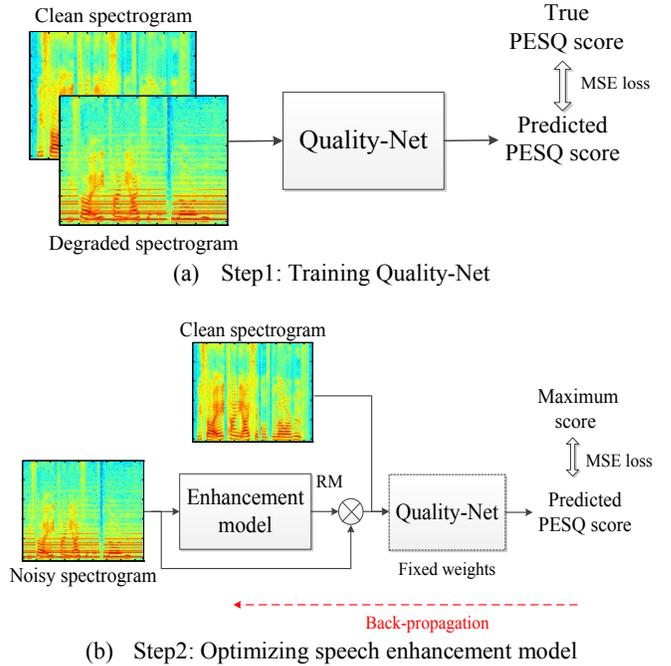

Fig. 1. Two steps of the proposed PESQ-maximization speech enhancement framework.

### B. Optimizing Enhancement Model with Fixed Quality-Net

Once Quality-Net is trained, it is concatenated at the output of a speech enhancement model. To train the enhancement model, the estimated quality scores are maximized while keeping the weights in Quality-Net fixed. In other words, Quality-Net is simply treated as a loss function that is highly correlated to the PESQ function. To prevent the enhancement model from generating additional artifacts in the enhanced spectrogram, its output is the ratio mask (RM) [28], which is a mask of value ranging between 0 to 1. Therefore, the optimal enhancement model $G^*$ can be obtained by solving the following optimization problem (we denote it as Quality-Net loss):

$$G^* = \arg\min_G \sum_{u=1}^{U} (1 - Q(N_u \otimes G(N_u), C_u))^2 \quad (1)$$

where $U$ is the total number of training utterances; $N_u$ and $C_u$ are the noisy and clean magnitude spectrograms of the $u$-th utterance, respectively. $Q$ represents Quality-Net and herein, we normalize the maximum value of the PESQ score to 1. $\otimes$ is the operator for element-wise multiplications. To obtain the time-domain waveform, the overlap–add method was applied using the enhanced magnitude spectrum with the noisy phase. The overall frameworks of these two steps are shown in Fig. 1. In this study, the enhancement model is pre-trained with the MSE loss first and then fine-tuned by the Quality-Net loss.

## III. Experiments

### A. TIMIT Dataset

In our experiments, the TIMIT corpus [29] was used to prepare the training, validation, and test sets. 300 utterances were randomly selected from the training set of the TIMIT database for training. These utterances were further corrupted with 10 noise types (crowd, 2 machine, alarm and siren, traffic and car, animal sound, water sound, wind, bell, and laugh noise) from [30], at five SNR levels (from -8 dB to 8 dB with steps of 4 dB) to form 15000 training utterances. To monitor the training process and choose proper hyperparameters, we randomly selected another 100 clean utterances from the TIMIT training set to form our validation set. Each utterance was further corrupted with one of the noise types (different from those already used in the training set) from [30] at five different SNR levels (from -10 dB to 10 dB with steps of 5 dB). To evaluate the performance of different training methods, 100 clean utterances from the TIMIT test set were randomly selected as our test set. These utterances were mixed with four unseen noise types (engine, white, street, and baby cry), at five SNR levels (-6 dB, 0 dB, 6 dB, 12 dB, and 18 dB). In summary, 2000 utterances were prepared to form the test set.

In addition to the noisy speech, the training set for Quality-Net also includes the enhanced speech by a BLSTM-based speech enhancement model (its structure is depicted in the next section), as in [27].

### B. Model Structure

The speech enhancement model proposed in this experiment is a BLSTM [31] model with two bidirectional LSTM layers, each with 200 nodes, followed by two fully connected layers, each with 300 LeakyReLU nodes and 257 sigmoid nodes for RM estimation, respectively. The parameters are trained with RMSprop, which is typically a suitable optimizer for RNNs.

Quality-Net herein is a CNN with four two-dimensional (2-D) convolutional layers with the number of filters and kernel size as follows: [15, (5, 5)], [25, (7, 7)], [40, (9, 9)], and [50, (11, 11)]. To handle the variable-length input, a 2-D global average pooling layer was added, so that the features were fixed with 50 dimensions (50 is the number of feature maps in the previous layer). Three fully connected layers were added subsequently, each with 50 and 10 LeakyReLU nodes, and 1 linear node, respectively. To make Quality-Net a smooth function (we do not want a small change in the input spectrogram can result in a significant difference to the estimated quality score), we constrained it to be 1-Lipschitz continuous by spectral normalization [32]. Our preliminary experiments on the validation set found that adding this constraint can yield a higher PESQ score to the proposed framework.

### C. Fine-tuning the Enhancement Model by Quality-Net Loss

In this section, we first demonstrate the relation between the training iterations and PESQ scores on the validation set. Experimental results show that the enhancement model trained with Quality-Net loss can increase the PESQ scores rapidly, and thus we report the "iteration" number (updated number of model after seeing a mini-batch of training data) instead of the epoch. Figure 2 shows the fine-tuning process of the conventional MSE loss and the proposed Quality-Net loss. Note that here the enhancement model was pre-trained by MSE loss with early stopping. As shown, training more iteration with MSE loss cannot further improve the score. On the other hand, Quality-Net loss can boost the performance within only a few iterations. This result implies that Quality-Net can extract essential speech quality information from the training data and incorporate such information in the model; thus, Quality-Net can provide instant and correct gradient directions when fine-tuning the enhancement model.

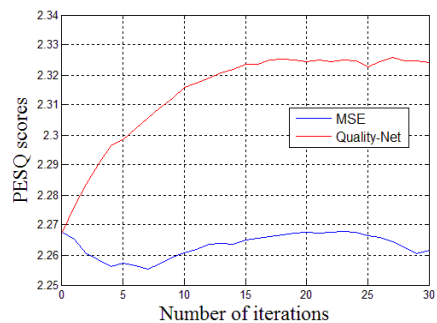

Fig. 2. Fine-tuning process of pre-trained enhancement model with different loss functions.

### D. Baselines

For the DNN baselines, there are three hidden layers with 256 rectifier linear units (ReLU) nodes. DNN (MSE) is trained to minimize the MSE between predicted and clean spectrogram. A stronger DNN baseline is based on the PESQ-inspired loss function, perceptual metric for speech quality evaluation (PMSQE), proposed by [26]. We also trained BLSTM models (same structure as depicted in section III-B) with different loss functions for comparison.

### E. Experimental Results

To verify the effectiveness of the proposed framework, the standard PESQ function was used to measure the speech quality and the score ranges from -0.5 to 4.5. We also presented STOI for speech intelligibility evaluation and the score ranges from 0 to 1. Both the two metrics are the higher the better. Table I presents the results of the average PESQ and STOI scores on the test set for the baselines and proposed method, which maximizes the score of Quality-Net (the loss functions are indicated in the parentheses). As shown in Table I, the DNN (PMSQE) performs much better than the DNN (MSE), and comparable to the BLSTM (MSE) in low SNR cases. Although BLSTM (PMSQE) can achieve the highest PESQ score in the low SNR conditions, its average STOI scores are worse than the BLSTM (MSE). When we pre-trained the BLSTM enhancement model with the MSE loss and subsequently fine-tuned by the Quality-Net loss, we could maintain the speech intelligibility with better speech quality (increase of 0.10 points) compared to the BLSTM baselines.

### F. Spectrogram Comparison

Next, we presented the spectrograms of a clean TIMIT utterance, the same utterance corrupted by street noise at 0 dB, enhanced speeches by BLSTM with MSE loss and fine-tuned by proposed Quality-Net loss in Fig. 3. From Fig. 3(c), we observed that although MSE loss can guide BLSTM effectively remove the background noise, some noise still remains. On the other hand, Fig. 3 (d) shows that the remaining noise can be further removed by the Quality-Net loss and hence the PESQ score improves. This agreed with the results shown in Table I that Quality-Net loss can boost the PESQ scores.

Table I
Performance comparisons of different models in terms of PESQ and STOI.

| SNR (dB) | Noisy | | DNN (MSE) | | DNN (PMSQE) [26] | | BLSTM (MSE) | | BLSTM (PMSQE)[26] | | Proposed BLSTM_pre-trained (Quality-Net) | |
|---|---|---|---|---|---|---|---|---|---|---|---|---|
| | PESQ | STOI | PESQ | STOI | PESQ | STOI | PESQ | STOI | PESQ | STOI | PESQ | STOI |
| 18 | 2.807 | 0.967 | 2.810 | 0.855 | 3.082 | 0.886 | 3.287 | **0.972** | 2.899 | 0.882 | **3.377** | 0.966 |
| 12 | 2.375 | 0.919 | 2.576 | 0.831 | 2.819 | 0.865 | 2.908 | **0.942** | 2.777 | 0.867 | **3.010** | 0.937 |
| 6 | 1.963 | 0.831 | 2.275 | 0.788 | 2.497 | 0.822 | 2.504 | **0.885** | 2.578 | 0.836 | **2.614** | 0.882 |
| 0 | 1.589 | 0.709 | 1.912 | 0.715 | 2.111 | 0.741 | 2.065 | **0.796** | **2.261** | 0.773 | 2.171 | 0.794 |
| -6 | 1.242 | 0.576 | 1.530 | 0.604 | 1.711 | 0.615 | 1.569 | 0.663 | **1.865** | **0.667** | 1.671 | 0.663 |
| Avg. | 1.995 | 0.800 | 2.221 | 0.759 | 2.444 | 0.786 | 2.467 | **0.852** | 2.476 | 0.805 | **2.569** | 0.848 |

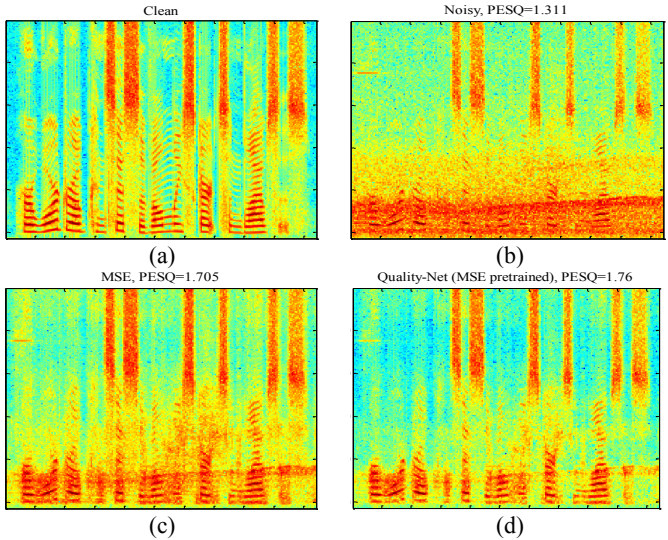

Fig. 3. Spectrograms of a TIMIT utterance: (a) clean speech, (b) noisy speech (street noise at 0 dB), (PESQ = 1.311), (c) enhanced speech by BLSTM with MSE loss (PESQ = 1.705) (d) enhanced speech by BLSTM (pre-trained) with Quality-Net loss (PESQ = 1.76).

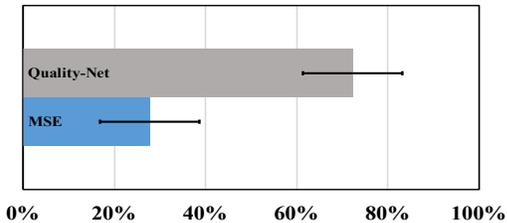

Fig. 4. Results of AB preference test (with 95% confidence intervals).

### G. Subjective evaluation

To evaluate the perceptual quality of the enhanced speech, we conducted AB preference tests to compare the proposed Quality-Net loss with the MSE loss. Each pair of samples are presented in a randomized order. For each listening test, 20 sample pairs were randomly selected from the test set; 15 listeners participated. Listeners were instructed to select the sample with the better quality. In Fig. 4, we can observe that Quality-Net loss significantly outperforms (with p-value 0.00027) MSE loss, without an overlap in the confidence intervals.

Table II
PESQ and STOI scores on the Voice Bank corpus.

| | PESQ | STOI |
|---|---|---|
| Noisy | 1.970 | 0.920 |
| Wiener filter | 2.223 | 0.914 |
| BLSTM (MSE) | 2.529 | **0.935** |
| **BLSTM (Quality-Net)** | **2.713** | 0.932 |

### H. Voice Bank corpus

In addition to the experiments tested on the TIMIT corpus, we also investigate the performance of BLSTM (MSE) and BLSTM (Quality-Net) on the noisy Voice Bank corpus [33, 34]. The results are listed in Table II. From the table, we can note consistent trends to those in Table I: the proposed BLSTM (Quality-Net) can further boost the PESQ score (increase of 0.184 points) and maintain the speech intelligibility.

### I. Discussion

In Fig. 2, we showed that gradients from Quality-Net can guide the enhancement model to further increase the PESQ scores. However, we also found that the gradient direction is correct only in the first few iterations. The true PESQ scores start to decrease (even though the predicted scores are still increasing) when the iteration number is large[1]. This is because the Quality-Net has not seen the speech generated by the updated enhancement model before. Therefore, Quality-Net is fooled [35] (estimated quality scores increase but true scores decrease) as the generation scheme of adversarial examples [36]. Solving this problem will be our future endeavor.

### IV. CONCLUSION

We herein proposed adopting Quality-Net as an approximated PESQ function to form the objective function for fine-tuning the speech enhancement models. This learned loss function successfully addressed the non-differentiable issue that was encountered during direct PESQ optimization. The experimental results indicated that minimizing Quality-Net loss can further significantly increase the PESQ scores. Because our method does not need to know any computational details of the interested metric, if a more advanced evaluation metric is proposed, our "black-box" method can be easily applied.

1: See: https://github.com/JasonSWFu/Learning-with-Learned-Loss-Function